\documentclass{amsproc}

\usepackage{amsbsy,amssymb,amscd,amsfonts,latexsym,amstext,delarray,
amsmath,epsfig}
\input xypic

\def\cancel#1#2{\ooalign{$\hfil#1\mkern1mu/\hfil$\crcr$#1#2$}}
\def\dirac{\mathpalette\cancel\partial}
\newcommand{\scr}{\mathcal}
\def\R{{\mathbb R}}
\def\H{{\mathbb H}}

\def\Z{{\mathbb Z}}
\def\Q{{\mathbb Q}}
\def\C{{\mathbb C}}

\def\N{{\mathbb N}}

\def\Ind{{\rm Ind}}
\def\Tr{{\rm Tr}}
\def\tr{{\rm tr}}

\def\cA{{\mathcal A}}

\def\cB{{\mathcal B}}
\def\cE{{\mathcal E}}
\def\cH{{\mathcal H}}

\def\cL{{\mathcal L}}

\def\cR{{\mathcal R}}
\def\cC{{\mathcal C}}

\def\cK{{\mathcal K}}

\def\cU{{\mathcal U}}

\newcommand{\ie}{{\it i.e.\/}\ }
\newcommand{\eg}{{\it e.g.\/}\ }
\newcommand{\cf}{{\it cf.\/}\ }

\newcommand{\M}{\scr{M}}

\newcommand{\cG}{\scr{G}}

\newcommand{\Spec}{{\rm Spec}}
\newcommand{\Proj}{{\rm Proj}}

\newcommand{\ind}{{\rm ind}}

\newcommand{\PSL}{{\rm PSL}}

\newtheorem{thm}{Theorem}[section]

\newtheorem{conj}{Conjecture}[section]

\numberwithin{equation}{section}
\title[FQHE: a noncommutative geometry perspective]{Towards the fractional
quantum Hall effect: a noncommutative geometry perspective}
\author{Matilde Marcolli}
\author{Varghese Mathai}
\address{M.~Marcolli: Max--Planck Institut f\"ur Mathematik  \\
Vivatsgasse 7 \\
Bonn, D-53111 Germany} \email{marcolli\@@mpim-bonn.mpg.de}
\address{V.~Mathai: Department of Pure Mathematics \\
University of Adelaide \\ 5005 Australia}
\email{vmathai\@@maths.adelaide.edu.au}

\begin{document}

\maketitle

In this paper we give a survey of some models of the integer
and fractional quantum
Hall effect based on noncommutative geometry. We begin by recalling
some classical geometry of electrons in solids and the passage to 
noncommutative geometry produced by the presence of a magnetic
field. We recall how one can obtain this way a single 
electron model of the integer quantum Hall effect. While in the case of the
integer quantum Hall effect the underlying geometry is Euclidean, we
then discuss a model of the fractional quantum Hall effect, which is based on hyperbolic 
geometry simulating
the multi-electron interactions. We derive the fractional 
values of the Hall conductance 
as integer multiples of orbifold Euler characteristics. 
We compare the results with experimental data.

\section{Electrons in solids -- Bloch theory and algebraic geometry}

We first recall some general facts about the mathematical theory
of electrons in solids. In particular, after reviewing some basic
facts about Bloch theory, we recall an approach pioneered by
Gieseker at al. \cite{Gies} \cite{Gies2}, which uses algebraic
geometry to treat the inverse problem of determining the
pseudopotential from the data of the electric and optical
properties of the solid.

\medskip
\subsection*{Crystals}%\hfill\medskip

The Bravais lattice of a crystal is a lattice $\Gamma \subset
\R^d$ (where we assume $d=2,3$), which describes the symmetries of
the crystal.

\smallskip

The electron--ions interaction is described by a periodic
potential
\begin{equation}\label{periodU}
U(x)= \sum_{\gamma\in \Gamma} u(x-\gamma),
\end{equation}
namely, $U$ is invariant under the translations in $\Gamma$,
\begin{equation}\label{transl}
T_\gamma U = U, \ \ \ \   \forall \gamma\in \Gamma.
\end{equation}

When one takes into account the mutual interaction of electrons,
one obtains an $N$-particles Hamiltonian of the form
\begin{equation}\label{Nparticle}
 \sum_{i=1}^N (-\Delta_{x_i}+U(x_i)) + \frac{1}{2} \sum_{i\neq j}
W(x_i-x_j).
\end{equation}
This can be treated in the {\em independent electron
approximation}, namely by introducing a modification $V$ of the
single electron potential
\begin{equation}\label{pseudopot}
 \sum_{i=1}^N (-\Delta_{x_i}+V(x_i)).
\end{equation}
It is remarkable that, even though the original potential $U$ is
unbounded, a reasonable independent electron approximation can be
obtained with $V$ a bounded function.

\smallskip

The wave function for the $N$-particle problem \eqref{pseudopot}
is then of the form $$\psi(x_1,\ldots, x_N)=\det(\psi_i(x_j)),$$
for $(-\Delta+V)\psi_i=E_i\psi_i$ so that $\sum
(-\Delta_{x_i}+V(x_i)) \psi =E \psi$ with $E=\sum E_i$. This
reduces a multi-electron problem to the single particle case.

\smallskip

However, in this approximation, usually the single electron
potential $V$ is not known explicitly, hence the focus shifts on
the {\em inverse problem} of determining $V$.

\medskip
\subsection*{Bloch electrons}%\hfill\medskip

Let $T_\gamma$ denote the unitary operator on $\cH=L^2(\R^d)$
implementing the translation by $\gamma\in \Gamma$, as in
\eqref{transl}. We have, for $H=-\Delta + V$,
\begin{equation}\label{Tgamma}
 T_\gamma \, H\, T_{\gamma^{-1}} = H, \ \ \ \ \forall \gamma \in
\Gamma.
\end{equation}
Thus, we can simultaneously diagonalize these operators. This can
be done via characters of $\Gamma$, or equivalently, via its
Pontrjagin dual $\hat\Gamma$. In fact, the eigenvalue equation is
of the form $T_\gamma \psi=c(\gamma)\psi$. Since
$T_{\gamma_1\gamma_2}=T_{\gamma_1}T_{\gamma_2}$, and the operators
are unitaries, we have $c:\Gamma \to U(1)$, of the form
$$
c(\gamma)=e^{i\langle k, \gamma\rangle}, \ \ \ k\in\hat\Gamma.
$$

\smallskip

The Pontrjagin dual $\hat\Gamma$ of the abelian group $\Gamma\cong
\Z^d$ is a compact group isomorphic to $T^d$, obtained by taking
the dual of $\R^d$ modulo the reciprocal lattice
\begin{equation}\label{recLat}
 \Gamma^\sharp = \{ k \in \R^d : \langle k, \gamma\rangle\in 2
\pi \Z, \forall \gamma \in \Gamma \}.
\end{equation}

\begin{center}
\begin{figure}
\includegraphics[scale=0.4]{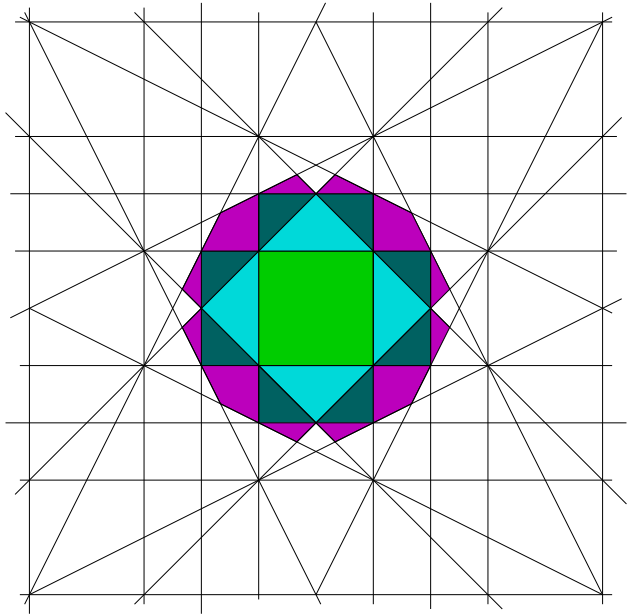}
\caption{Brillouin zones in a 2D crystal}
\end{figure}
\end{center}

\medskip
\subsection*{Brillouin zones}%\hfill\medskip

By definition, the Brillouin zones of a crystals are fundamental
domains for the reciprocal lattice $\Gamma^\sharp$ obtained via
the following inductive procedure. The {\em Bragg hyperplanes} of
a crystal are the hyperplanes along which a pattern of diffraction
of maximal intensity is observed when a beam of radiation (X-rays
for instance) is shone at the crystal. The $N$-th Brillouin zone
consists of all the points in (the dual) $\R^d$ such that the line
from that point to the origin crosses exactly $(n-1)$ Bragg
hyperplanes of the crystal.

\medskip
\subsection*{Band structure}%\hfill\medskip

One obtains this way (\cf \cite{Gies}) a family self-adjoint
elliptic boundary value problems, parameterized by the lattice
momenta $k\in \R^d$,
\begin{equation}\label{ellDk}
 D_k = \left\{ \begin{array}{cc} (-\Delta +V) \psi = E \psi \\
\psi(x+\gamma)=e^{i\langle k, \gamma\rangle} \psi(x)
\end{array} \right.
\end{equation}
For each value of the momentum $k$, one has eigenvalues $\{
E_1(k), E_2(k), E_3(k), \ldots \}$. As functions of $k$, these
satisfy the periodicity
$$ E(k)=E(k+u) \ \ \ \forall u\in \Gamma^\sharp. $$
It is customary therefore to plot the eigenvalue $E_n(k)$ over the
n-the Brillouin zone and obtain this way a map
$$ k \mapsto E(k) \ \ \ k\in \R^d $$
called the {\em energy--crystal momentum dispersion relation}.

\medskip
\subsection*{Fermi surfaces and complex geometry}%\hfill\medskip

\begin{center}
\begin{figure}
\includegraphics[scale=0.65]{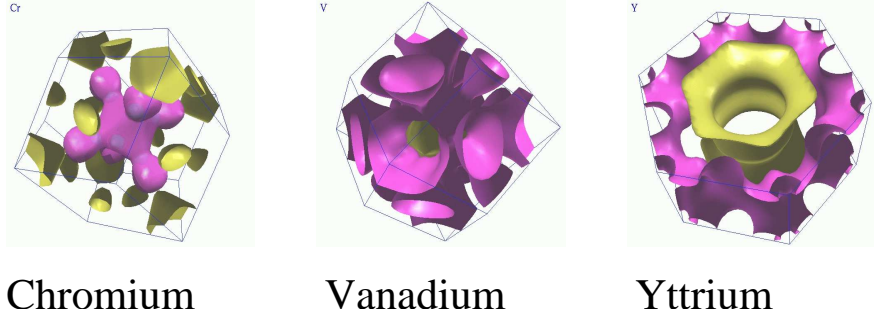}
\caption{Examples of Fermi surfaces \label{FigFS}}
\end{figure}
\end{center}

Many electric and optical properties of the solid can be read off
the geometry of the Fermi surface. This is a hypersurface $F$ in
the space of crystal momenta $k$,
\begin{equation}\label{Fermisurf}
 F_\lambda(\R) = \{ k\in \R^d : E(k)=\lambda \}.
\end{equation}
A comprehensive archive of Fermi surfaces for various chemical
elements can be found in \cite{FSdata}, or online at {\tt
http://www.phys.ufl.edu/fermisurface}. We reproduce in Figure
\ref{FigFS} an example of the complicated geometry of Fermi
surfaces.

The approach to the theory of electrons in solids proposed by
\cite{Gies} \cite{Gies2} is based on the idea that the geometry of
the Fermi surfaces can be better understood by passing to complex
geometry and realizing \eqref{Fermisurf} as a cycle on a complex
hypersurface. One considers first the {\em complex Bloch variety}
defined by the condition
\begin{equation}\label{Blochvar}
 B(V)=\left\{ (k,\lambda)\in \C^{d+1} : \begin{array}{cc} \exists
\psi \ \
\text{nontrivial solution of} \\
 (-\Delta +V) \psi = \lambda \psi \\
\psi(x+\gamma)=e^{i\langle k, \gamma\rangle} \psi(x)
\end{array} \right\}
\end{equation}
Then the complex Fermi surfaces are given by the fibers of the
projection to $\lambda\in \C$,
\begin{equation}\label{CFS}
F_\lambda(\C)=\pi^{-1}(\lambda) \subset B(V).
\end{equation}
This is a complex hypersurface in $\C^d$. To apply the tools of
projective algebraic geometry, one works in fact with a singular
projective hypersurface (a compactification of $B(V)$, \cf
\cite{Gies}).

One can then realize the original Fermi surface \eqref{Fermisurf}
as a cycle $F_\lambda \cap \R^d =F_\lambda(\R) $ representing a
class in the homology $H_{d-1}(F_\lambda(\C),\Z)$. A result of
\cite{Gies} is that the {\em integrated density of states}
\begin{equation}\label{intdens}
\rho(\lambda)=\lim_{\ell \to \infty} \frac{1}{\ell} \# \{
\text{eigenv of } H \leq \lambda \},
\end{equation}
for $H=-\Delta+V$ on $L^2(\R^d/\ell\Gamma)$, is obtained from a
{\em period}
\begin{equation}\label{period}
 \frac{d\rho}{d\lambda} = \int_{F_\lambda(\R)} \omega_\lambda,
\end{equation}
where $\omega_\lambda$ is a holomorphic differential on
$F_\lambda(\C)$.

\medskip
\subsection*{Discretization}%\hfill\medskip
It is often convenient to treat problems like \eqref{ellDk} by
passing to a discretized model. On $\ell^2(\Z^d)$, one considers
the {\em random walk} operator
\begin{equation}\label{randomwalk}
 \begin{array}{ll} \cR \psi(n_1,\ldots, n_d) = &
+\sum_{i=1}^d \psi(n_1,\ldots,n_i+1,\ldots, n_d)\\[2mm]
& +\sum_{i=1}^d \psi(n_1,\ldots,n_i-1,\ldots, n_d). \end{array}
\end{equation}
This is related to the {\em discretized Laplacian} by
\begin{equation}\label{dLapl}
 \Delta \psi(n_1,\ldots, n_d) =  (2d -\cR)\,\, \psi(n_1,\ldots,
n_d).
\end{equation}

\smallskip

In this discretization, the complex Bloch variety is described by
a polynomial equation in $z_i,z_i^{-1}$ (\cf \cite{Gies})
$$
B(V)=\left\{(z_1\ldots,z_d,\lambda): \begin{array}{l} \exists
\psi\in \ell^2(\Gamma)
\text{ nontriv sol of } \\[2mm] (\cR+V)\psi =(\lambda+2d)\,\psi \\[2mm]
R_{\gamma_i}\psi = z_i \,\psi\end{array}\right\},
$$
where $R_{\gamma_i}\psi(n_1,\ldots, n_d)
=\psi(n_1,\ldots,n_i+a_i,\ldots, n_d)$.

\smallskip

It will be very useful for us in the following to also consider a
more general random walk for a discrete group $\Gamma$, as an
operator on $\cH=\ell^2(\Gamma)$. Let $\gamma_i$ be a symmetric
set of generators of $\Gamma$, \ie a set of generators and their
inverses. The random walk operator is defined as
\begin{equation}\label{randomGamma}
 \cR\, \psi (\gamma) = \sum_{i=1}^r R_{\gamma_i} \psi\, (\gamma)
= \sum_{i=1}^r \psi (\gamma \gamma_i)
\end{equation}
and the corresponding discretized Laplacian is $\Delta=r-\cR$.

\medskip
\subsection*{The breakdown of classical Bloch theory}%\hfill\medskip

The approach to the study of electrons in solids via Bloch theory
breaks down when either a magnetic field is present, or when the
periodicity of the lattice is replaced by an aperiodic
configuration, such as those arising in quasi--crystals. What is
common to both cases is that the commutation relation $T_\gamma H
=H T_\gamma$ fails.

\smallskip

Both cases can be studied by replacing ordinary geometry by {\em
noncommutative geometry} (\cf \cite{Co}). Ordinary Bloch theory is
replaced by noncommutative Bloch theory \cite{Gruber}. A good
introduction to the treatment via noncommutative geometry of the
case of aperiodic solids can be found in \cite{BellC}.

\smallskip

For our purposes, we are mostly interested in the other case,
namely the presence of magnetic field, as that is the source of
the Hall effects. Bellissard pioneered an approach to the quantum
Hall effect via noncommutative geometry and derived a complete and
detailed mathematical model for the Integer Quantum Hall Effect
within this framework, \cite{Bell}.

\section{Quantum Hall Effect}

We describe the main aspects of the classical and quantum (integer
and fractional) Hall effects, and some of the current approaches
used to produce a mathematical model. Our introduction will not be
exhaustive. In fact, for reasons of space, we will not discuss
many interesting mathematical results on the quantum Hall effect
and will concentrate mostly on the direction leading to the use of
noncommutative geometry.

\medskip
\subsection*{Classical Hall effect}%\hfill\medskip

The classical Hall effect was first observed in the XIX century
\cite{Hall}. A thin metal sample is immersed in a constant uniform
strong orthogonal magnetic field, and a constant current ${\bf j}$
flows through the sample, say, in the $x$-direction. By Flemming's
rule, an electric field is created in the $y$-direction, as the
flow of charge carriers in the metal is subject to a Lorentz force
perpendicular to the current and the magnetic field. This is
called the {\em Hall current}.

\smallskip

The equation for the equilibrium of forces in the sample
\begin{equation} \label{clHall} N e {\bf E} + {\bf j} \wedge {\bf B}
=0, \end{equation} defines a linear relation. The ratio of the
intensity of the Hall current to the intensity of the electric
field is the Hall conductance,
\begin{equation} \label{condHall} \sigma_H = \frac{N e \delta}{B}.
\end{equation}
In the stationary state, $\sigma_H$ is proportional to the
dimensionless {\em filling factor} $\nu = \frac{\rho h}{eB}$,
where $\rho$ is the 2-dimensional density of charge carriers, $h$
is the Planck constant, and $e$ is the electron charge. More
precisely, we have
\begin{equation} \label{linearHall} \sigma_H=\frac{\nu}{R_H},
\end{equation}
where $R_H= h/e^2$ denotes the Hall resistance, which is a
universal constant. This measures the fraction of Landau level
filled by conducting electrons in the sample.

\begin{figure}
\begin{center}
\includegraphics{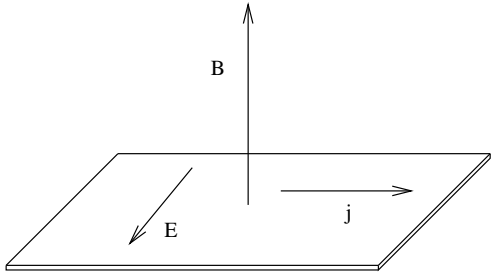}
\caption{Hall effect}
\end{center}
\end{figure}

\medskip
\subsection*{Integer quantum Hall effect}%\hfill\medskip

In 1980, von Klitzing's experiment showed that, lowering the
temperature below 1 K, quantum effects dominate, and the relation
of Hall conductance to filling factor shows plateaux at integer
values, \cite{vKl}. The effect is measured with very high
precision (of the order of $10^{-8}$) and allows for a very
accurate measurement of the fine structure constant $e^2/\hbar c$.

\smallskip

Under the above conditions, one can effectively use the
independent elector approximation discussed in the previous
section and reduce the problem to a single particle case.

\smallskip

The main physical properties of the integer quantum Hall effect
are the following:
\begin{itemize}
\item $\sigma_H$, as a function of $\nu$, has plateaux at integer
multiples of $e^2/h$.
\item At values of $\nu$ corresponding to the plateaux, the conductivity
along the current density axis (direct conductivity) vanishes.
\end{itemize}

\smallskip

Laughlin first suggested that IQHE should have a geometric
explanation \cite{La}. More precisely, the fact that the
quantization of the Hall conductance appears as a very robust
phenomenon, insensitive to changes in the sample and its geometry,
or to the presence of impurities, suggests the fact that the
effect should have the same qualities of the {\em index theorem},
which assigns an integer invariant to an elliptic differential
operator, in a way that is topological and independent of
perturbations. The prototype of such index theorems is the
Gauss--Bonnet theorem, which extracts from an infinitesimally
variable quantity, the curvature of a closed surface, a robust
topological invariant, its Euler characteristic. The idea of
modelling the integer quantum Hall effect on an index theorem
started fairly early after the discovery of the effect. Laughlin's
formulation can already be seen as a form of Gauss--Bonnet, while
this was formalized more precisely in such terms shortly
afterwards by Thouless et al. (1982) and by Avron, Seiler, and
Simon (1983) (\cf \cite{Thou} \cite{ASS}).

\smallskip

One of the early successes of Connes' noncommutative geometry
\cite{Co} was a rigorous mathematical model of the integer quantum
Hall effect, developed by Bellissard, van Elst, and Schulz-Baldes,
\cite{Bell}. Unlike the previous models, this accounts for all the
aspects of the phenomenon: integer quantization, localization,
insensitivity to the presence of disorder, and vanishing of direct
conductivity at plateaux levels. Again the integer quantization is
reduced to an index theorem, albeit of a more sophisticated
nature, involving the Connes--Chern character, the $K$-theory of
$C^*$-algebras and cyclic cohomology (\cf \cite{Connes}).

\medskip
\subsection*{Fractional quantum Hall effect}%\hfill\medskip

\begin{center}
\begin{figure}
\includegraphics[scale=0.6]{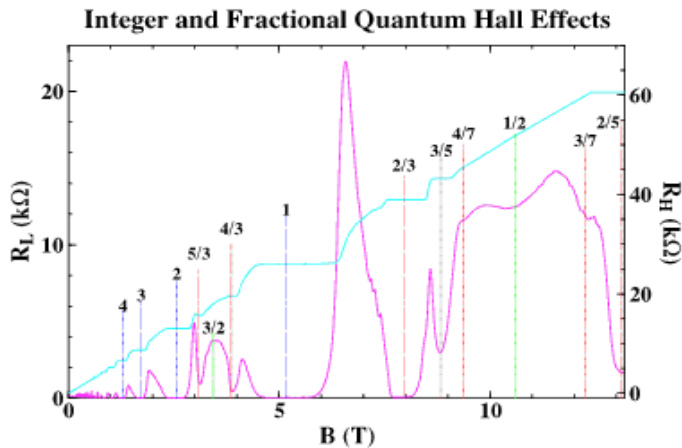}
\caption{Fractional quantum Hall effect \label{Figfqhe}}
\end{figure}
\end{center}

The fractional QHE was discovered by Stormer and Tsui in 1982. The
setup is as in the quantum Hall effect: in a high quality
semi-conductor interface, which will be modelled by an infinite
2-dimensional surface, with low carrier concentration and
extremely low temperatures $\sim 10 mK$, in the presence of a very
strong magnetic field, the experiment shows that the same graph of
$\frac{h}{e^2}\sigma_H$ against the filling factor $\nu$ exhibits
plateaux at certain fractional values (Figure \ref{Figfqhe}).

Under the conditions of the experiments, the independent electron
approximation that reduces the problem to a single electron is no
longer viable and one has to incorporate the Coulomb interaction
between the electrons in a many-electron theory. For this reason,
many of the proposed mathematical models of the fractional quantum
Hall effect resort to quantum field theory and, in particular,
Chern--Simons theory (\cf \eg \cite{RGPhi}).

\smallskip

In this survey we will only discuss a proposed model \cite{MM1}
\cite{MM2}, which is based on extending the validity of the
Bellissard approach to the setting of hyperbolic geometry as in
\cite{CHMM}, where passing to a negatively curved geometry is used
as a device to simulate the many-electrons Coulomb interaction
while remaining within a single electron model.

\smallskip

What is expected of any proposed mathematical model? Primarily
three things: to account for the strong electron interactions, to
exhibit the observed fractions and predict new fractions, and to
account for the varying width of the observed plateaux. We will
discuss these various aspects in the rest of the paper.

\section{Noncommutative geometry models}

In the theory of the quantum Hall effect noncommutativity arises
from the presence of the magnetic field, which has the effect of
turning the Brillouin zone into a noncommutative space. In
Bellissard's model of the integer quantum Hall effect \cite{Bell}
the noncommutative space obtained this way is the noncommutative
torus and the integer values of the Hall conductance are obtained
from the corresponding Connes--Chern character. We will consider a
larger class of noncommutative spaces, associated to the action of
a Fuchsian group of the first kind without parabolic elements on
the hyperbolic plane. The idea is that, by effect of the strong
interaction with the other electrons, a single electron ``sees''
the surrounding geometry as curved, while the lattice sites appear
to the moving electron, in a sort of multiple image effect, as
sites in a lattice in the hyperbolic plane. This model will
recover the integer values but will also produce fractional values
of the Hall conductance.

\medskip
\subsection{Hyperbolic geometry}%\hfill\medskip

Let $\H$ denote the hyperbolic plane. Its geometry is described as
follows. Consider the pseudosphere $\{ x^2+y^2+z^2 -t^2 =1 \}$ in
4-dimensional Minkowski space-time $M$. The $z=0$ slice of the
pseudosphere realizes an isometric embedding of the hyperbolic
plane $\H$ in $M$. In this geometry, a periodic lattice on the
resulting surface is determined by a Fuchsian group $\Gamma$ of
isometries of $\H$ of signature $(g; \nu_1, \ldots, \nu_n)$,
\begin{equation}\label{FuchsianG}
\Gamma = \Gamma(g; \nu_1, \ldots, \nu_n).
\end{equation}
This is a discrete cocompact subgroup $\Gamma\subset \PSL(2,\R)$
with generators $a_i, b_i, c_j$, with $i=1,\ldots,g$ and
$j=1,\ldots,n$ and a presentation of the form
\begin{equation}\label{presentation}
\Gamma(g; \nu_1, \ldots, \nu_n)=\langle a_i, b_i, c_j\,\,
\left|\,\, \prod_{i=1}^g [a_i,b_i]c_1\cdots c_n =1, \,\,\,
c_j^{\nu_j} =1 \right. \rangle.
\end{equation}

\begin{center}
\begin{figure}
\includegraphics[scale=0.5]{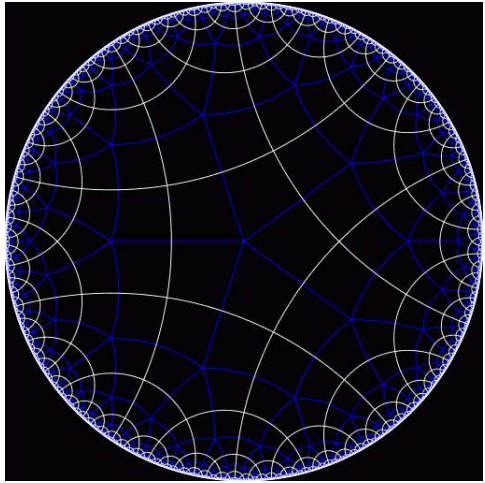}
\caption{Tiling of the hyperbolic plane \label{Figtile}}
\end{figure}
\end{center}

The quotient of the action of $\Gamma$ by isometrieson $\H$,
\begin{equation}\label{hyporb}
\Sigma(g; \nu_1, \ldots, \nu_n):= \Gamma \backslash \H,
\end{equation}
is a hyperbolic orbifold, namely a compact Riemann surface of
genus $g$ with $n$ cone points $\{ x_1, \ldots, x_n \}$, which are
the image of points in $\H$ with non-trivial stabilizer of the
action of $\Gamma$. In the torsion free case, where we only have
generators $a_i$ and $b_i$, we obtain smooth compact Riemann
surfaces of genus $g$.

\medskip
\subsection*{Orbifolds}%\hfill\medskip

\begin{center}
\begin{figure}
\includegraphics[scale=0.6]{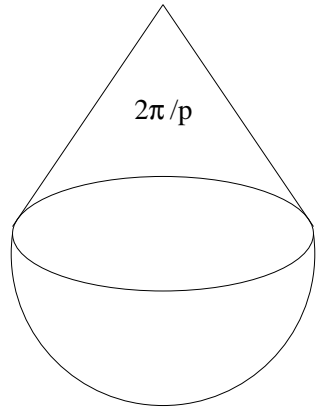}
\caption{Thurston's teardrop orbifold \label{Figtear}}
\end{figure}
\end{center}

The space $\Sigma(g; \nu_1, \ldots, \nu_n)$ of \eqref{hyporb} is a
special case of {\em good orbifolds}. These are orbifolds that are
orbifold--covered by a smooth manifold. In dimension two, in the
oriented compact case, the only exceptions (bad orbifolds) are the
Thurston teardrop (Figure \ref{Figtear}) with a single cone point
of angle $2\pi/p$, and the double teardrop.

In particular, all the hyperbolic orbifolds $\Sigma(g; \nu_1,
\ldots, \nu_n)$ are good orbifolds and they are orbifold--covered
by a smooth compact Riemann surface,
\begin{equation}\label{orbcover}
\Sigma_{g'} \stackrel{G}{\longrightarrow}
\Sigma(g;\nu_1,\ldots,\nu_n)=\Gamma\backslash \H,
\end{equation}
where the genus $g'$ satisfies the Riemann--Hurwitz formula for
branched covers
\begin{equation}\label{genuscount}
g'=1 +\frac{\# G}{2} (2(g-1)+(n-\nu)),
\end{equation}
for $\nu=\sum_{j=1}^n \nu_j^{-1}$.

\smallskip

Notice moreover that the orbifolds $\Sigma=\Sigma(g; \nu_1,
\ldots, \nu_n)$ are an example of classifying space for proper
actions in the sense of Baum--Connes, namely they are of the form
\begin{equation}\label{BGproper}
\Sigma=\underline{B}\Gamma=\Gamma\backslash\underline{E}\Gamma.
\end{equation}

\smallskip

An important invariant of orbifold geometry, which will play a
crucial role in our model of the fractional quantum Hall effect,
is the {\em orbifold Euler characteristic}. This is an analog of
the usual topological Euler characteristic, but it takes values in
rational numbers, $\chi_{orb}(\Sigma)\in \Q$. It is multiplicative
over orbifold covers, it agrees with the usual topological Euler
characteristic $\chi$ for smooth manifolds, and it satisfies the
inclusion--exclusion principle
\begin{equation}\label{inclexcl}
 \begin{array}{rl}
\chi_{orb}(\Sigma_1\cup\cdots\cup\Sigma_r)= & \sum_i
\chi_{orb}(\Sigma_i) - \sum_{i,j}\chi_{orb}(\Sigma_i\cap\Sigma_j)\\[2mm]
 \cdots & + (-1)^{r+1}
\chi_{orb}(\Sigma_1\cap\cdots\cap\Sigma_r) .\end{array}
\end{equation}

\smallskip

In the case of the hyperbolic orbifolds
$\Sigma(g;\nu_1,\ldots,\nu_n)$, the orbifold Euler characteristic
is given by the formula
\begin{equation}\label{orbEulch}
\chi_{orb}(\Sigma(g;\nu_1,\ldots,\nu_n))=2-2g +\nu -n .
\end{equation}

\medskip
\subsection*{Magnetic field and symmetries}%\hfill\medskip

The magnetic field can be described by a 2-form $\omega =d\eta$,
where $\omega$ and $\eta$ are the field and potential,
respectively, subject to the customary relation ${\bf B}={\rm
curl} {\bf A}$.

\smallskip

One then considers the magnetic Schr\"odinger operator
\begin{equation}\label{magnschrod}
\Delta^\eta + V,
\end{equation}
where the magnetic Laplacian is given by $\Delta^\eta:=(d
-i\eta)^*\,\, (d-i\eta)$ and $V$ is the electric potential of the
independent electron approximation.

\smallskip

The 2-form $\omega$ satisfies the periodicity condition
$\gamma^*\omega = \omega$, for all $\gamma\in \Gamma=\Z^d$ (\eg
one might assume that the magnetic field is a constant field $B$
perpendicular to the sample). Thus, we have the relation
$0=\omega-\gamma^*\omega =d( \eta-\gamma^*\eta )$, which implies
\begin{equation}\label{etaper}
 \eta-\gamma^*\eta  = d \phi_\gamma.
\end{equation}

\smallskip

Due to the fact that $\eta$ itself need not be periodic, but only
subject to condition \eqref{etaper}, the magnetic Laplacian no
longer commutes with the $\Gamma$ action, unlike the ordinary
Laplacian. This is, in a nutshell, how turning on a magnetic field
brings about noncommutativity.

\smallskip

What are then the symmetries of the magnetic Laplacian? These are
given by the {\em magnetic translations}. Namely, after writing
\eqref{etaper} in the form $\phi_\gamma(x)=\int_{x_0}^x
(\eta-\gamma^*\eta)$, we consider the unitary operators
\begin{equation}\label{magntransl}
 T_\gamma^\phi \psi := \exp(i\phi_\gamma)\,\,
T_\gamma \psi.
\end{equation}
It is easy to check that these satisfy the desired commutativity
$(d -i\eta)T_\gamma^\phi = T_\gamma^\phi (d -i\eta)$. However,
commutativity is still lost in another way, namely, magnetic
translations, unlike the ordinary translations by elements
$\gamma\in \Gamma=\Z^d$, no longer commute among themselves
(except in the case of integer flux). We have instead
\begin{equation}\label{mtcomm}
  T_\gamma^\phi T_{\gamma'}^\phi =\sigma(\gamma,\gamma')
T^\phi_{\gamma\gamma'}.
\end{equation}

\smallskip

Instead of obtaining a representation of $\Gamma$, the magnetic
translations give rise to a {\em projective representation}, with
the cocycle
\begin{equation}\label{sigma}
 \sigma(\gamma,\gamma')=\exp(-i\phi_\gamma(\gamma' x_0)),
\end{equation}
where $\phi_\gamma(x)+\phi_{\gamma'}(\gamma x)
-\phi_{\gamma'\gamma}(x)$ is independent of $x$.

\medskip
\subsection*{Algebra of observables}

The $C^*$-algebra of observables should be minimal, yet large enough to 
contain all of the spectral projections
onto gaps in the spectrum of the magnetic Schr\"odinger operators
$\Delta^\eta + V$ for  any periodic potential $V$. Now let $\cU$ denote 
the set of all bounded operators on $L^2(\H)$ that commute with the 
magnetic translations. By a theorem of von Neumann, $\cU$ is a von Neumann
algebra. By Lemma 1.1, \cite{KMS}, any element $Q \in \cU$ can be represented
uniquely as $$Q = \sum_{\gamma\in \Gamma} T^{-\phi}_\gamma \otimes Q(\gamma),$$
where $Q(\gamma)$ is a bounded operator on the Hilbert space $L^2(\H/\Gamma)$. 
Let $\cL^1$ denote the subset of $\cU$ consisting of all bounded operators on 
 $Q$  on  $L^2(\H)$  that commute with the 
magnetic translations and such that 
$\sum_{\gamma\in \Gamma} || Q(\gamma)|| < \infty$. The norm closure of 
$\cL^1$ is a $C^*$-algebra denoted by $\cC^*$, that is taken to be the algebra
of observables. Using the Riesz representation for projections
onto spectral gaps cf.\eqref{holproj}, one can show as in \cite{KMS} that $C^*$ contains 
all of the projections onto the spectral gaps of the magnetic  
Schr\"odinger operators. In fact, it can be shown that $C^*$ is Morita
equivalent to the reduced twisted group $C^*$ algebra
$C^*_r(\Gamma, \bar\sigma)$,  explained later in the text, showing
that in both the continuous and the discrete models for the quantum Hall effect,
the algebra of observables are Morita equivalent, so they describe the same physics.
Hence we will mainly discuss the discrete model in this paper. 

\medskip
\subsection*{Semiclassical properties, as the electro-magnetic 
coupling constant goes to zero}

Recall the magnetic Schr\"odinger operator
\begin{equation}
\Delta^\eta + \mu^{-2} V,
\end{equation}
where $\Delta^\eta$ is the magnetic Laplacian, $V$ is the electric potential 
and $\mu$ is the  electro-magnetic coupling constant. When $V$ is a Morse
type potential, i.e. for all $x\in \H$, $V(x) \ge 0$. Moreover, if  $V(x_0)=0$
for some $x_0$ in $\M$, then  there
is a positive constant $c$ such that $V(x)\ge c|x-x_0|^2 I$ for
all $x$ in a neighborhood of $x_0$.
Also assume that $V$ has at least one zero point. Observe that all
 functions $V=|df|^2$, where $|df|$ denotes the
pointwise norm of the differential of a $\Gamma$-invariant Morse
function $f$ on $\H$, are examples of Morse type potentials.

Under these assumptions, the semiclassical properties of the 
spectrum of  the magnetic Schr\"odinger operator, and the Hall conductance
were studied by Kordyukov, Mathai and Shubin in \cite{KMS}, as the electro-magnetic coupling constant $\mu$ goes to zero.
One result obtained is that there exists an arbitrarily large number of gaps in the spectrum
of the magnetic Schr\"odinger operator for all $\mu$ sufficiently small. Another result obtained
in \cite{KMS} is that the low energy bands do not contribute to the Hall conductance, 
 again for all $\mu$ sufficiently small.

\medskip
\subsection*{Extending Pontrjagin duality}%\hfill\medskip

An advantage of noncommutative geometry is that it provides a
natural generalization of Pontrjagin duality. Namely, the duals of
discrete groups are noncommutative spaces.

\smallskip

In fact, first recall that, if $\Gamma$ is a discrete abelian
group, then its Pontrjagin dual $\hat\Gamma$, which is the group
of characters of $\Gamma$ is a compact abelian group. The duality
is given by Fourier transform $e^{i\langle k, \gamma\rangle}$, for
$\gamma\in \Gamma$ and $k\in \hat\Gamma$.

\smallskip

In particular, this shows that the algebra of functions on
$\hat\Gamma$ can be identified with the (reduced) $C^*$-algebra of
the group $\Gamma$,
\begin{equation}\label{CGammadual}
C(\hat\Gamma) \cong C^*_r(\Gamma),
\end{equation}
where the reduced $C^*$-algebra $C^*_r(\Gamma)$ is the
$C^*$-algebra generated by $\Gamma$ in the regular representation
on $\ell^2(\Gamma)$.

\smallskip

When $\Gamma$ is non-abelian, although Pontrjagin duality no
longer works in the classical sense, the left hand side of
\eqref{CGammadual} still makes perfect sense and behaves ``like''
the algebra of functions on the dual group. In other words, we can
say that, for a non-abelian group, the Pontrjagin dual
$\hat\Gamma$ still exists as a noncommutative space whose algebra
of functions is $C^*_r(\Gamma)$.

\smallskip

This point of view can be adopted to work with the theory of
electrons in solids whenever classical Bloch theory breaks down.
In the case of aperiodicity, the dual $\hat\Gamma$ (which is
identified with the Brillouin zone) is replaced by a
noncommutative $C^*$-algebra. This is the case, similarly, for the
presence of magnetic field in the quantum Hall effect. The
magnetic field deforms the Brillouin zone to a noncommutative
space, given by the (noncommutative) algebra of the magnetic
translation.

\medskip
\subsection*{Harper operators}%\hfill\medskip

It is again convenient to discretize the problem. The discretized
magnetic Laplacian is given in terms of the Harper operator, which
is an analog of the random walk operator seen in the previous
section, but defined using the magnetic translations. For
$\Gamma=\Z^2$, the Harper operator is of the form

\begin{equation}\label{2dHarper}
 \begin{array}{rll} H_{\alpha_1,\alpha_2}\psi(m,n)=  &
e^{-i\alpha_1 n}
&\psi(m+1,n)\\[2mm] + &  e^{i\alpha_1 n} &\psi(m-1,n) \\[2mm]
+ & e^{-i\alpha_2 m} &\psi(m,n+1)\\[2mm] + & e^{i\alpha_2 m} &\psi
(m,n-1).
\end{array}
\end{equation}
Here the 2-cocycle $\sigma:\Gamma\times\Gamma \to U(1)$ is given
by
\begin{equation}\label{sigmaZ2}
\sigma((m',n'),(m,n))=\exp(-i(\alpha_1 m'n +\alpha_2 m n')).
\end{equation}
The magnetic translations are generated by $U=R_{(0,1)}^\sigma$
and $V=R_{(1,0)}^\sigma$ of the form
\begin{equation}\label{UVgen}
 U \psi(m,n)= \psi(m,n+1) e^{-i\alpha_2 m}  \ \ \ \
 V \psi(m,n) = \psi(m+1,n) e^{-i\alpha_1 n}.
\end{equation}
These satisfy the commutation relations of the {\em noncommutative
torus} $\cA_\theta$, with $\theta = \alpha_2-\alpha_1$, namely
\begin{equation}\label{NCtorus}
UV = e^{i\theta} VU.
\end{equation}
The Harper operator \eqref{2dHarper} is in fact more simply
written as $H_\sigma = U + U^* + V+ V^* $.

\begin{figure}
\begin{center}
\includegraphics{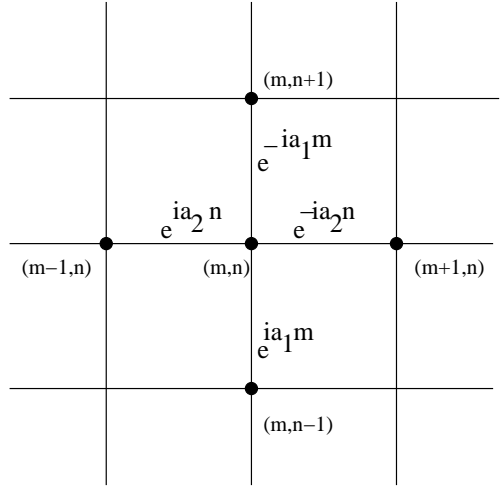}
\caption{{\small The Harper operator on the square
lattice}}
\end{center}
\end{figure}

\smallskip

This shows that, on a 2-dimensional lattice, the effect of the
magnetic field is to deform the usual Brillouin zone (which is an
ordinary torus $T^2$) to a noncommutative torus, where the
parameter $\theta$ depends on the magnetic flux through a cell of
the lattice.

\smallskip

As in the case of the discretization of the ordinary Laplacian,
for the magnetic Laplacian we can also consider the corresponding
Harper operator on a more general (possibly non-abelian) discrete
group $\Gamma$. This will be useful later, in our model of the
fractional quantum Hall effect, but we introduce it here for
convenience. For the general setup for finitely generated discrete
groups recalled here below, we follow \cite{CHMM}.

\smallskip

Suppose given a finitely generated discrete group $\Gamma$ and a
{\em multiplier} $\sigma: \Gamma \times \Gamma \to U(1)$ (a
2-cocycle)
$$ \sigma(\gamma_1,\gamma_2)\sigma(\gamma_1\gamma_2,\gamma_3)=
\sigma(\gamma_1,\gamma_2\gamma_3)\sigma(\gamma_2,\gamma_3), $$
$\sigma(\gamma,1)=\sigma(1,\gamma)=1$.

\smallskip

On the Hilbert space $\ell^2(\Gamma)$, consider the {\em
left/right $\sigma$-regular representations}
\begin{equation}\label{leftright}
 L^\sigma_\gamma \psi (\gamma')= \psi(\gamma^{-1} \gamma')
\sigma(\gamma, \gamma^{-1}\gamma')  \ \ \ \ \  R^\sigma_\gamma
\psi(\gamma') = \psi(\gamma' \gamma) \sigma(\gamma', \gamma).
\end{equation}
These satisfy
\begin{equation}\label{LRcomm}
 L^\sigma_\gamma L^\sigma_{\gamma'} = \sigma(\gamma,\gamma')
L^\sigma_{\gamma\gamma'} \ \ \ \ \ \  R^\sigma_\gamma
R^\sigma_{\gamma'} = \sigma(\gamma,\gamma')
R^\sigma_{\gamma\gamma'}.
\end{equation}
The cocycle identity can be used to show that the left
$\sigma$-regular representation commutes with the right
$\bar{\sigma}$-regular representation, where $\bar{\sigma}$
denotes the conjugate cocycle.  Also the left
$\bar{\sigma}$-regular representation commutes with the right
$\sigma$-regular representation.

\smallskip

Let $\{\gamma_i\}_{i=1}^r$ be a symmetric set of generators of
$\Gamma$. The Harper operator is given by
\begin{equation}\label{HarperGamma}
 \cR_\sigma = \sum_{i=1}^r  R^\sigma_{\gamma_i}.
\end{equation}
The operator $r-\cR_\sigma$ is the discrete analog of the magnetic
Laplacian (\cf \cite{Sun}).

\begin{figure}
\begin{center}
\includegraphics{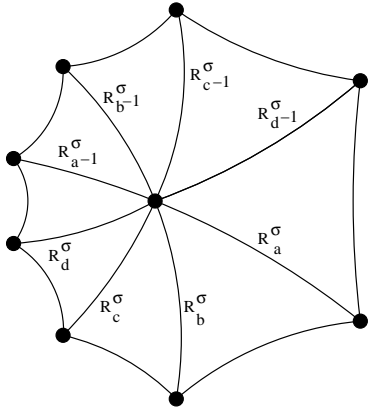}
\caption{Harper operator on a lattice in the hyperbolic plane}
\end{center}
\end{figure}

\medskip
\subsection*{Algebra of observables (discrete model)}%\hfill\medskip

We continue in the same generality as above. The special case of
interest for the integer quantum Hall effect will be for
$\Gamma=\Z^2$, but we adopt a more general setting in view of
applications to the fractional case.

\smallskip

For $\Gamma$ a finitely generated discrete group, let
$\C(\Gamma,\sigma)$ be the algebra generated by the magnetic
translations represented as operators in $\cB(\ell^2(\Gamma))$
through the right $\sigma$-regular representation
$R^\sigma_\gamma$. Equivalently, the algebra $\C(\Gamma,\sigma)$
consists of functions
$$ f:\Gamma \to \C $$
with the convolution product
$$f_1* f_2(\gamma) = \sum_{\gamma_1\gamma_2=\gamma} f_1(\gamma_1)
f_2(\gamma_2) \sigma(\gamma_1, \gamma_2), $$ acting on the Hilbert
space $\ell^2(\Gamma)$.

\smallskip

By taking the weak closure of $\C(\Gamma,\sigma)$ one obtains the
{\em twisted group von Neumann algebra} $\cU(\Gamma,\sigma)$. This
is equivalently (by the commutant theorem of von Neumann)
described as
\[
\cU(\Gamma,\sigma)=\left\{A\in B(\ell^2(\Gamma)):
[L_{\gamma}^{\bar\sigma}, A]=0\; \forall \gamma\in \Gamma\right\}.
\]
That is, $\cU(\Gamma,\sigma)$ is the commutant of the left
$\bar{\sigma}$-regular representation. When taking the norm
closure of $\C(\Gamma,\sigma)$ one obtains the twisted (reduced)
group $C^*$-algebra $C^*_r(\Gamma,\sigma)$, which is the algebra
of observables in the discrete model.

\smallskip

The key properties of these algebras are summarized as follows.
$\cU(\Gamma,\sigma)$ is generated by its projections and it is
also closed under the {\em measurable} functional calculus, \ie if
$a\in\cU(\Gamma,\sigma)$ and $a=a^*$, $a>0$, then $f(a)\in
\cU(\Gamma,\sigma)$ for all {\em essentially bounded measurable}
functions $f$ on $\R$. On the other hand, $C^*_r(\Gamma,\sigma)$
has only at most countably many projections and is only closed
under the  {\em continuous} functional calculus.

\smallskip

In the case when $\sigma=1$ (integer flux), with the group $\Gamma
=\Z^2$, we simply have $\cU(\Gamma,1) \cong L^\infty(T^2)$ and
$C^*(\Gamma,1) \cong C(T^2)$, \ie functions on the classical
Brillouin zone. Here the ordinary torus $T^2$ is identified with 
the group $\widehat \Gamma$ of characters
of the abelian group $\Gamma = \Z^2$.

\smallskip

In Bellissard's model of the integer quantum Hall effect, where
$\Gamma=\Z^2$, with $\sigma$ the nontrivial cocycle described in
\eqref{sigmaZ2} and $\theta = \alpha_2-\alpha_1$, the twisted
(reduced) group $C^*$-algebra is the irrational rotation algebra
of the noncommutative torus,
\begin{equation}\label{tgCtorus}
  C^*_r(\Gamma,\sigma) \cong A_\theta.
\end{equation}

\smallskip

We will not describe in detail the derivation of the quantization
of the Hall conductance in this model of the integer quantum Hall
effect. In fact, we will concentrate mostly on a model for the
fractional quantum Hall effect and we will show how to recover the
integer quantization within that model, using the results of
\cite{CHMM}.

\medskip
\subsection*{Spectral theory}%\hfill\medskip

For $\Gamma$ a finitely generated discrete group and
$\{g_i\}_{i=1}^r$ a symmetric set of generators, the {\em Cayley
graph} $\cG=\cG(\Gamma, g_i)$ has as set of vertices the elements
of $\Gamma$ and as set of edges emanating from a given vertex
$h\in \Gamma$ the set of translates $g_i h$.

\smallskip

The random walk operator \eqref{randomGamma} for $\Gamma$ is then
an average on nearest neighbors in the Cayley graph. The discrete
analog of the Schr\"odinger equation is of the form
\begin{equation}\label{schroedGamma}
 i\frac{\partial}{\partial t}\psi=\cR_\sigma \psi +V\psi,
\end{equation}
where all physical constants have been set equal to 1. It
describes the quantum mechanics of a single electron confined to
move along the Cayley graph of $\Gamma$, subject to the periodic
magnetic field. Here $\cR_\sigma$ is 
the Harper operator encoding 
the magnetic field and $V$ is the electric potential of the
independent electron approximation. The latter can be taken to be
an operator in the twisted group algebra, $V\in \C(\Gamma,
\sigma)$. More precisely, equation \eqref{schroedGamma} should
be formulated with the discrete magnetic laplacian
$\delta_{\sigma}= r - \cR_\sigma$ in place of $\cR_\sigma$, 
with $r$ the cardinality of a symmetric
set of generators for $\Gamma$. This does not really matter as far as 
the spectral properties are concerned, as the spectrum of one 
determines the spectrum of the other.  

\smallskip

As in the case of the theory of electrons in solids without
magnetic field recalled in the first section, an important problem
is understanding the energy levels of the Hamiltonian
$H_{\sigma,V}=\cR_\sigma + V$, and the band structure (gaps in the
spectrum).

\smallskip

The Harper operator $\cR_\sigma$ is a bounded {\em self-adjoint}
operator on $\ell^2(\Gamma)$, since it is defined in terms of a
{\em symmetric} set of generators of $\Gamma$. Thus, the spectrum
$\Spec(\cR_{\sigma})$ is a closed and bounded subset of $\R$. It
follows that the complement $\R\backslash \Spec(\cR_{\sigma})$  is
an open subset of $\R$, hence a countable union of disjoint open
intervals. Each such interval is called a {\em gap in  the
spectrum}.

\smallskip

\begin{center}
\begin{figure}
\includegraphics[scale=0.6]{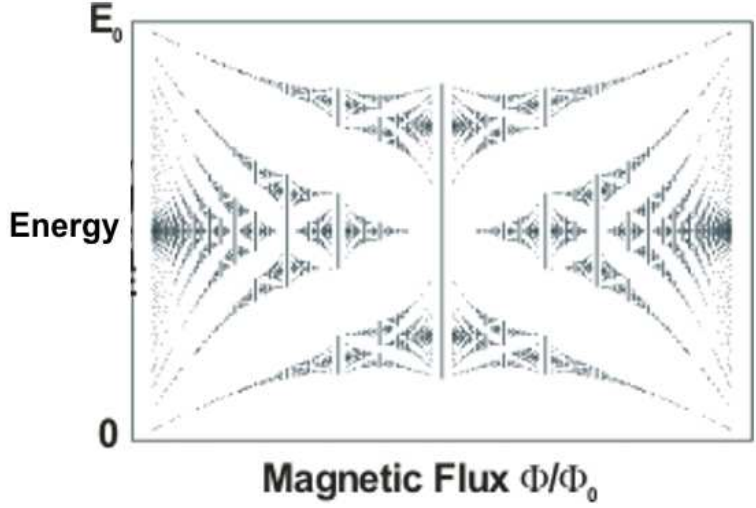}
\caption{Hofstadter butterfly \label{FigHoft}}
\end{figure}
\end{center}

There are two very different situations. When the complement of
the spectrum consists of a finite collection of intervals then the
operator has a band structure, while if the complement consists of
an infinite collection of intervals then the spectrum is a Cantor
set. In the case of the group $\Gamma=\Z^2$, one or the other
possibility occurs depending on the rationality or irrationality
of the flux
$$ \theta = \langle [\sigma],[\Gamma] \rangle. $$
This gives rise to a diagram known as the Hofstadter butterfly
(Figure \ref{FigHoft}).

\medskip
\subsection*{Range of the trace}%\hfill\medskip

In our model of the fractional quantum Hall effect, $\Gamma$ is a
cocompact Fuchsian group of signature $\;(g,  \nu_1,\ldots,
\nu_n)$. In this case (\cf \cite{MM2}), if $[\sigma]$ is {\em
rational}, then there is only a finite number of gaps in the
spectrum of $H_{\sigma}+V$. In fact, if $\theta=p/q$  then the
number of gaps is at most
\begin{equation}\label{count}
(q+1)\prod_{j=1}^n (\nu_j +1).
\end{equation}

\smallskip

In terms of the algebra of observables, the question of how many
gaps there are in the spectrum of $H_{\sigma, V}$ can be reduced
to studying the {\em number of projections} in the $C^*$-algebra
$C_r^*(\Gamma,\sigma)$ (up to equivalence). In fact, we have
\[ H_{\sigma, V}\in\C (\Gamma,\sigma)\subset
C_r^*(\Gamma,\sigma)\subset\cU(\Gamma,\sigma). \] In particular,
$H_{\sigma}$ and its spectral projections
\[ P_{E}=\chi_{(-\infty,E]}(H_{\sigma, V}) \]
belong to the algebra $\cU(\Gamma,\sigma)$. Moreover, when
$E\notin\Spec(H_{\sigma, V})$, the spectral projection $P_E$ is in
$C_r^*(\Gamma,\sigma)$. In fact, suppose that the spectrum of
$H_{\sigma, V}$ is contained in a closed interval, and that the
open interval $(a,b)$ is a spectral gap of $H_{\sigma, V}$.
Suppose that $E\in (a,b)$, \ie $E\notin\Spec(H_{\sigma, V})$. Then
there is a holomorphic function $\phi$ on a neighborhood of ${\rm
spec}(H_{\sigma, V})$ such that
\begin{equation}\label{holproj}
P_E=\phi(H_{\sigma, V}) = \int_C \frac{d\lambda}{\lambda -
H_{\sigma, V}}
\end{equation}
where $C$ is a closed contour enclosing the spectrum of
$H_{\sigma,V}$ to the left of $E$. Since $C_r^*(\Gamma,\sigma)$ is
closed under the holomorphic functional calculus, it follows that
$P_E\in C_r^*(\Gamma,\sigma)$.

\smallskip

The equivalence relation we need to consider on projections, so
that the counting will provide the counting of spectral gaps, is
described as follows. Let $\Proj(C_r^*(\Gamma,\sigma)\otimes \cK)$
denote the projections in $C_r^*(\Gamma,\sigma)\otimes  \cK$,
where $\cK$ the $C^*$ algebra of compact operators. Two
projections $P, Q \in \Proj(C_r^*(\Gamma,\sigma)\otimes\cK)$ are
said to be {\em Murray-von Neumann equivalent} if there is an
element $V\in C_r^*(\Gamma,\sigma)\otimes \cK$ such that $P= V^*V$
and $Q= VV^*$, and we write $P\sim Q$. It can be shown that
$\Proj(C_r^*(\Gamma,\sigma)\otimes\cK)/\sim$ is an abelian
semi-group under direct  sums, and  the Grothendieck group
$K_0(C_r^*(\Gamma,\sigma)))$ is defined as the associated abelian
group.

\smallskip

Now the estimate on the number of equivalence classes of
projections is achieved by computing the range of a trace. The von
Neumann algebra $\cU(\Gamma,\sigma)$ and $C^*$-algebra
$C_r^*(\Gamma,\sigma)$ have a canonical faithful finite trace
$\tau$, where
$$ \tau(a) = \langle  a\delta_1, \delta_1\rangle_{\ell^2(\Gamma)}, $$
where $\delta_\gamma$ is the basis of $\ell^2(\Gamma)$. If $\Tr$
denotes the standard trace on bounded operators in an
$\infty$-dimensional separable Hilbert space $\cH$, then we obtain
a trace
\[ \tr=\tau\otimes\Tr:\; \Proj(C_r^*(\Gamma,\sigma)\otimes\cK)\to \R. \]
This induces a trace on the $K$-group
\[ [\tr]: K_0(C_r^*(\Gamma,\sigma)))\to \R \]
with
\[ \tr(\Proj(C_r^*(\Gamma,\sigma)))=[\tr](K_0(C_r^*(\Gamma,\sigma)))\cap [0,1]. \]
The result quoted above in \eqref{count}, counting the energy gaps
in our hyperbolic model, can then be derived from the following
result proved in \cite{MM2}.

\begin{thm}
Let $\Gamma$ be a cocompact Fuchsian group of signature $(g:
\nu_1, \ldots, \nu_n)$ and $\sigma$ be a multiplier on $\Gamma$
with flux $\theta$. Then the range of the trace is,
\begin{equation}\label{trrange}
[\tr](K_0(C_r^*(\Gamma,\sigma)))=\Z+\theta\Z + \sum_j
\frac{1}{\nu_j}\Z.
\end{equation}
\end{thm}

Here the flux is again given by the pairing
$\theta=\langle[\sigma],[\Gamma]\rangle$, where
$[\Gamma]=\frac{[\Sigma_{g'}]}{\# G}$ is the fundamental class of
the group $\Gamma=\Gamma(g;\nu_1,\ldots,\nu_n)$ and $g'$ is given
by the formula \eqref{genuscount}.

\smallskip

The Baum--Connes conjecture holds for the Fuchsian groups
$\Gamma=\Gamma(g;\nu_1,\ldots,\nu_n)$, and one can compute the
$K$-theory of the $C^*$-algebra $C_r^*(\Gamma,\sigma)$ in terms of
the orbifold K-theory of $\Sigma(g;\nu_1,\ldots,\nu_n)$. This uses
a Morita equivalence $(A\otimes C_0(G))\rtimes \Gamma \simeq
C_0(\Gamma\backslash G,\cE)$, where $\cE=A\times_\Gamma G \to
\Gamma\backslash G$, in the case where $G=\PSL(2,\R)$. Without the
twisting by $\sigma$, one can identify
$$ K_\bullet(C_r^*(\Gamma))  \cong
K^\bullet_{SO(2)}(P(g;\nu_1,\ldots,\nu_n)), $$ where
$P(g;\nu_1,\ldots,\nu_n)$ is the frame bundle $\Gamma\backslash
\PSL(2,\R)$. The result can be identified with the orbifold
K-theory
$$ K^\bullet_{orb}(\Sigma(g;\nu_1,\ldots,\nu_n))  \cong
\left\{ \begin{array}{ll}  \Z^{2-n+\sum\nu_j} & \bullet=even \\
\Z^{2g} & \bullet=odd \end{array}\right. $$ In the twisted case,
one still has the equivalence $C_0(\Gamma\backslash G,\cE)\simeq
C_0(\Gamma\backslash G,\cE_\sigma)$ when the class
$\delta(\sigma)=0$, where $\delta: H^2(\Gamma,U(1)) \to
H^3(\Gamma,\Z)$ is a surjection coming from the long exact
sequence of $1\to\Z\hookrightarrow \R\stackrel{\exp(2\pi
i\cdot)}{\longrightarrow} U(1) \to 1$.

\smallskip

The computation of the range of the trace \eqref{trrange} then
follows from an index theorem. Let $\cE$ be an orbifold vector
bundle over $\Sigma=\Sigma(g;\nu_1,\ldots,\nu_n)$, and $[\cE]\in
K^\bullet_{orb}(\Sigma)$. Let $\tilde\dirac_\cE^+$ be the twisted
Dirac operator on the universal cover $\H$. For $\nabla^2=i\omega$
the magnetic field, the operator $\tilde\dirac_\cE^+\otimes\nabla$
commutes with the projective action of $(\Gamma,\sigma)$. There is
an analytic index
\begin{equation}\label{indK}
\ind_{(\Gamma,\sigma)} \tilde\dirac_\cE^+\otimes\nabla \in
K_0(C^*_r(\Gamma,\sigma)),
\end{equation}
which is the image under the (twisted) Kasparov map
$$ \mu_\sigma([\cE]) = \ind_{(\Gamma,\sigma)}
\tilde\dirac_\cE^+\otimes\nabla. $$

\smallskip

To compute the range of the trace $[\tr]:
K_0(C^*_r(\Gamma,\sigma)) \to \R$ one computes then the index
\begin{equation}\label{indL2}
\Ind_{L^2}\left(\tilde\dirac_\cE^+\otimes\nabla\right) =
[\tr]\left(\ind_{(\Gamma,\sigma)}
\tilde\dirac_\cE^+\otimes\nabla\right).
\end{equation}
We have
$$ \Ind_{L^2}\left(\tilde\dirac_\cE^+\otimes\nabla\right) =
\frac{1}{2\pi} \int_\Sigma \hat A \,\tr(e^{R^\cE}) e^\omega. $$
Since $\Sigma$ is of real dimension 2, this formula reduces to
$$ \frac{{\rm rank} \cE}{2\pi} \int_\Sigma \omega + \frac{1}{2\pi}
\int_\Sigma \tr(R^\cE). $$ The first term is computed by
$$ \frac{{\rm rank} (\cE)}{2\pi} \int_\Sigma \omega =
\frac{{\rm rank} (\cE)}{2\pi \# G} \int_{\Sigma_{g'}} \omega =
{\rm rank}( \cE) \langle [\sigma],[\Gamma]\rangle \in \theta \Z,$$
while the second term is computed by the Kawasaki index theorem
for orbifolds
$$ \Z \ni \ind(\dirac_\cE^+) =\frac{1}{2\pi}
\int_\Sigma \tr(R^\cE) + \frac{1}{2\pi} \sum_{i=1}^n
\frac{\beta_i}{\nu_i}, $$ where $(\beta_i,\nu_i)$ are the Seifert
invariants of $\cE$. This implies that
$$ \frac{1}{2\pi} \int_\Sigma \tr(R^\cE) \in \Z + \sum_{i=1}^n
\frac{1}{\nu_i}\Z. $$

\medskip

Rieffel, and Pimsner and Voiculescu established analogous results
in the case $\Gamma = \Z^2$. The result in the case of
torsion-free Fuchsian groups was established in \cite{CHMM}. In
more recent work, Mathai generalized this result to discrete
subgroups of rank 1 groups and to all amenable groups, and more
generally whenever the Baum--Connes conjecture with coefficients
holds for the discrete group, \cite{Mathai}. By contrast, the
behavior of spectral gaps when the flux is irrational is still
mysterious. The problem can be formulated in terms of the
following conjecture (also known as the ``generalized ten Martini
problem''), \cf \cite{MM1} \cite{MM2}.

\begin{conj}
Let $\Gamma$ be a cocompact Fuchsian group and $\sigma$ be a
multiplier on $\Gamma$. If the flux $\theta$ is irrational, then
there is a $V \in \mathbb C(\Gamma, \sigma)$ such that $H_{\sigma,
V}$ has an infinite number of gaps in its spectrum.
\end{conj}

It is not yet known if {\it any} gaps exist at all in this case!
However, using Morse--type potentials, Mathai and Shubin
\cite{MaShu} proved that there is an arbitrarily large number of
gaps in the spectrum of magnetic Schr\"odinger operators on
covering spaces, (\ie in the continuous model).

\smallskip

Recent work of Dodziuk, Mathai, and Yates \cite{MDY} shows another
interesting property of the spectrum, namely the fact that all
$L^2$ eigenvalues of the Harper operators of surface groups
$\Gamma$ are {\em algebraic numbers}, whenever the multiplier is
algebraic, that is, when $[\sigma] \in H^2(\Gamma,
\overline{\Q}/\Z)$. In fact the same result remains true when
adding potentials $V$ in $\overline{\Q}(\Gamma, \sigma)$ to the
Harper operator.

\section{Hall conductance}

We finally come to a discussion of the quantization of the Hall
conductance. This will follow again from a topological argument,
and index theorem, as in the Bellissard case, but in our setting
with hyperbolic geometry. We will derive, from our model, a
formula for the Hall conductance in terms of values of the
orbifold Euler characteristic, and we will compare the results
with experimentally observed values.

\medskip
\subsection*{A smooth subalgebra}%\hfill\medskip

We will consider a cyclic cocycle associated to the Connes--Kubo
formula for the conductance, which will be defined in terms of
certain derivations. For this reason, we need to introduce a
smooth subalgebra, namely, a dense involutive subalgebra of the
algebra of observables $C^*_r(\Gamma,\sigma)$. This subalgebra
contains $\C(\Gamma, \sigma)$ and is contained in the domain of
definition of the derivations. It contains the spectral projection
$P_E$, when the Fermi level is in a gap of the energy spectrum.
Moreover, it satisfies the following two key properties.
\begin{enumerate}
\item The inclusion $\cR \subset C^*_r(\Gamma,\sigma)$
induces an isomorphism in $K$-theory.
\item Polynomial growth group cocycles on $\Gamma$ define
cyclic cocycles on $\C(\Gamma, \sigma)$ that extend continuously
to $\cR$.
\end{enumerate}

\smallskip

$\cR$ is defined as follows.  Consider an operator $D$ defined as
$$ D \delta_\gamma = \ell(\gamma)  \delta_\gamma\;\forall \gamma \in
\Gamma, $$ where $\ell(\gamma)$ denotes the word length of
$\gamma$. Let $\delta = {\rm ad}(D)$ denote the commutator $[D,
\cdot]$. Then $\delta$ is an unbounded, but closed derivation on
$C^*_r(\Gamma,\sigma)$. Define
$$\cR := \bigcap_{k\in \N}{\rm Dom}(\delta^k).$$
It is clear that $\cR$ contains $\delta_\gamma\;\forall \gamma \in
\Gamma$ and so it contains $\C(\Gamma, \sigma)$. Hence it is dense
in $C^*_r(\Gamma,\sigma)$. It is not hard to see that $\cR$ is
closed under the holomorphic functional calculus, and therefore by
a result of Connes, property $(1)$ above holds, and by equation
\eqref{holproj}, $P_E \in \cR$.

\smallskip

Until now, we have not used any special property of the group
$\Gamma$. But now assume that $\Gamma$ is a surface group. Then it
follows from a result of \cite{Ji}, \cite{KMS} that there is a $k\in \N$ and
a positive constant $C'$ such that for all $f\in \mathbb C(\Gamma,
\sigma)$, one has the {\em Haagerup inequality}
\begin{equation}\label{Haagerup}
\| f\| \le C'\; \nu_k(f),
\end{equation}
where $\| f \|$ denotes the operator norm of the operator on
$\ell^2(\Gamma)$ given by left convolution by $f$, and the bound
$\nu_k(f)$ is given in terms of the $L^2$ norms of $f$ and of
$(1+l^2)^{s/2} f$, for all $0 \le s \le k$. Using this, it is
routine to show that property $(2)$ holds.

\smallskip

Notice that the spectral projections onto gaps in the Hamiltonian
$H$ belong to the algebra of observables $\cR$, for any choice of
electric potential $V$.

\medskip
\subsection*{Cyclic cocycles}%\hfill\medskip

Cyclic cohomology was introduced by Connes in \cite{Connes}. It is
a main source of invariants of noncommutative spaces, obtained by
the pairing of cyclic cocycles with $K$-theory. Cyclic cocycles
are also called multilinear traces, and the word cyclic refers to
invariance under the cyclic group $\Z/(n+1)\Z$ acting on the slots
of the Cartesian product. Namely, $t$ is a cyclic $n$-cocycle if
$$
 t : \cR\times \cR \cdots
\times \cR \to \C
$$
satisfies the cyclic condition
$$
t(a_0, a_1, \ldots, a_n) = t(a_n, a_0, a_1, \ldots, a_{n-1}) =
\cdots = t(a_1, \ldots, a_n, a_0),
$$
and the cocycle condition
%\begin{gather*}
$$
t(a a_0, a_1, \ldots, a_n) - t(a, a_0 a_1, \ldots, a_n) \cdots %\\
(-1)^{n+1} t(a_n a, a_0, \ldots, a_{n-1}) =0.
$$
%\end{gather*}

\smallskip

For instance, a {\em cyclic 0-cocycle} is just a trace. In fact,
in this case, the condition it satisfies is $t(ab) = t(ba)$. A
{\em cyclic 1-cocycle} satisfies $t(a, b) = t(b, a)$ and $t(ab, c)
- t(a, bc) + t(ca, b) = 0$, and a {\em cyclic 2-cocycle} satisfies
$$ t(a,b,c) = t(c,a,b) = t(b, c, a) \ \ \text{ and } $$
$$
t(ab, c, d) - t(a, bc, d) + t(a, b, cd) - t(da, b, c) = 0.
$$

\medskip
\subsection*{Conductance cocycle}%\hfill\medskip

A formula for the Hall conductance is obtained from transport
theory. In the case of $\Gamma=\Z^2$, the current density in $e_1$
direction corresponds to the functional derivative $\delta_1$ of
$H_\sigma$ by $A_1$, the corresponding component of the magnetic
potential. The expected value of current is the given by $\tr(P
\delta_1 H)$ for a state $P$ of the system. Using $\partial_t
P=i[P,H]$ and $\partial_t=\frac{\partial A_2}{\partial t} \times
\delta_2$, where $e_2\perp e_1$, one gets
$$ i\tr(P[\partial_t P,\delta_1 P]) = -i E_2 \tr(P[\delta_2 P,\delta_1
P]), $$ where the electrostatic potential has been gauged away,
leaving ${\bf E}=-\frac{\partial {\bf A}}{\partial t}$. Because
the charge carriers are Fermions, two different charge carriers
must occupy different quantum eigenstates of the Hamiltonian $H$.
In the zero temperature limit, charge carriers occupy all levels
below the Fermi level, so that we can set $P=P_F$ in the formula
above. This gives the Kubo formula for the conductance
$$ \sigma_H = \tr( P_F [\delta_1 P_F,\delta_2 P_F]). $$

\smallskip

This argument can be generalized to our setting, keeping into
account the fact that, in our model, by effect of the strong
multi-electron interaction, to a moving elector the directions $\{
e_1,e_2 \}$ appear split into $\{ e_i,e_{i+g} \}_{i=1,\ldots,g}$
corresponding to $a_i,b_i$, for some lattice in the hyperbolic
plane. The following is a general mathematical formulation of the
result.

\smallskip

Given a 1-cocycle $a$ on the discrete group  $\Gamma$, \ie
$$
a(\gamma_1\gamma_2) = a(\gamma_1) + a(\gamma_2) \qquad \forall
\gamma_1,\gamma_2\in \Gamma ,
$$
one can define a linear functional $\delta_a$ on the twisted group
algebra $\C(\Gamma, \sigma)$
$$
\delta_a (f) (\gamma) = a(\gamma) f(\gamma) .
$$
Then one verifies that $\delta_a$ is a derivation:
$$
\begin{array}{l}
\delta_a (fg) (\gamma) = a(\gamma) fg(\gamma)\\[2mm]
 = a(\gamma) \sum_{\gamma = \gamma_1\gamma_2} f(\gamma_1) g(\gamma_2)
\sigma(\gamma_1,\gamma_2)\\[2mm]
 = \sum_{\gamma = \gamma_1\gamma_2} \Big(a(\gamma_1) +
a(\gamma_2)\Big) f(\gamma_1) g(\gamma_2)
\sigma(\gamma_1,\gamma_2)\\[2mm]
 = \sum_{\gamma = \gamma_1\gamma_2} \Big(\delta_a (f)(\gamma_1)
g(\gamma_2) \sigma(\gamma_1,\gamma_2) + f(\gamma_1) \delta_a (g)
(\gamma_2)
\sigma(\gamma_1,\gamma_2)\Big)\\[2mm]
 = (\delta_a (f) g) (\gamma) + (f \delta_a g) (\gamma).
\end{array} $$

\smallskip

In the case of a Fuchsian group $\Gamma$, the first cohomology
$H^1(\Gamma, \Z)$ of the group $\Gamma$ is a free Abelian group of
rank $2g$, where $g$ is the genus of $\Gamma\backslash\H$. The
cohomology $H^1(\Gamma, \R)$ is in fact a symplectic vector space,
and we can assume that $\{ a_j, b_j\}_{j=1,\ldots, g}$ is a
symplectic basis.

\smallskip

We denote $\delta_{a_j}$ by $\delta_j$ and $\delta_{b_j}$ by
$\delta_{j+g}$. These derivations give rise to a cyclic 2-cocycle
on the twisted group algebra $\C(\Gamma, \sigma)$,
\begin{equation}\label{trK}
 \tr_K(f_0, f_1, f_2) =
\sum_{j=1}^g \tr(f_0  (\delta_j(f_1)\delta_{j+g}(f_2) -
  \delta_{j+g}(f_1)\delta_j(f_2))).
\end{equation}
$\tr_K$ is called the {\em conductance 2-cocycle}.

\smallskip

Let $P_E$ denote denote the spectral projection associated to the
Fermi level, \ie $P_E = \chi_{(-\infty, E]}(H)$. Then, in the zero
temperature limit, the Hall conductance is given by
$$
\sigma_E = \tr_K (P_E, P_E, P_E).
$$

\medskip
\subsection*{Quantum adiabatic limit}%\hfill\medskip

We recall briefly the justification of \eqref{trK} in terms of the
quantum adiabatic limit for a slowly varying time dependent
Hamiltonian, \cf \cite{CHM2}.

\smallskip

If $H(s)$ is a smooth family of self-adjoint Hamiltonians and
$P(s)$ are spectral projections on a gap in the spectrum of
$H(s)$, then
$$ X(s)= \frac{1}{2\pi i} \oint_C R(z,s) \partial_s P(s) R(z,s) dz, $$
with $R(z,s) = (H(s) - z)^{-1}$, satisfies the commutation
relations
$$ [\partial_s P(s), P(s)]=[H(s),X(s)]. $$

\smallskip

The quantum adiabatic limit theorem (\cf \cite{ASY}) then shows
that the adiabatic evolution approximates well the physical
evolution, for large values of the adiabatic parameter $\tau \to
\infty$, via an estimate of the form
$$ \| (U_\tau(s)-U_a(s))P(0) \| \leq $$
$$ \frac{1}{\tau} \max_{s\in [0,\infty)} \{ 2 \| X(s) P(s)\| + \|
\partial_s (X(s) P(s)) P(s) \| \}. $$
Here the physical evolution satisfies
$$ i\partial_s U_\tau (s)= \tau H (s) U_\tau (s), $$
$U_\tau(0)=1$, where $s=t/\tau$ is a scaled time, and the
adiabatic evolution is defined by the equation
$$ P(s) = U_a(s) P(0) U_a(s)^* $$
with $U_a(0)=1$.

\smallskip

In our setting, the functional derivative $\delta_k H$, with
respect to a component $A_k$ of the magnetic potential, gives a
current density $J_k$. Its expectation value in a state described
by a projection $P$ on a gap in the spectrum of the Hamiltonian is
then computed by $\tr(P\delta_k H)$. In the quantum adiabatic
limit, one can replace $\delta_k H$ with $\delta_k H_a$, where the
adiabatic Hamiltonian $H_a$ satisfies
$$ i \partial_s U_a(s) = \tau H_a(s)U_a(s) $$
and the equation of motion
$$ [H_a(s),P(s)]=\frac{i}{\tau} \partial_s P(s). $$
This implies that the relation
\begin{equation}\label{variation1}
 \tr (P[\partial_t P, \delta_k P]) = i \tr(\delta_k (P H_a)) - i \tr
(P \delta_k H_a).
\end{equation}
We make some simplifying assumptions. If the trace is invariant
under variations of $A_k$, then the first term in the right hand
side of \eqref{variation1} vanishes. We also assume that the only
time dependence of $H$ and $P$ is in the adiabatic variation of a
component $A_j$ distinct from $A_k$, and we work in the Landau
gauge, so that the electrostatic potential vanishes and the
electric field is given by ${\bf E}=-\partial {\bf A}/\partial t$.
Then we have $\partial_t = -E_j \delta_j$, so that the expectation
of the current $J_k$ is given by
$$ \tr(P\delta_k H) = i\tr(P[\partial_t P, \delta_k P]) $$
$$ = -i E_j \tr(P[\delta_j P, \delta_k P]), $$
hence the conductance for a current in the $k$ direction induced
by an electric field in the $j$ direction is given by $-i
\tr(P[\delta_j P, \delta_k P])$. The analytic aspects of this
formal argument can be made rigorous following the techniques used
in \cite{Xia}.

\medskip
\subsection*{Area cocycle}%\hfill\medskip

Our conclusion above, as in the case of the integer Hall effect,
is that one can compute the Hall conductance by evaluating a
certain cyclic cocycle on a projection, namely on some element in
K-theory. It is often the case that, in order to compute the
pairing of a cyclic cocycle with K-theory, one can simplify the
problem by passing to another cocycle in the same cohomology
class, \ie that differs by a coboundary. This is what will happen
in our case.

\smallskip

We introduce another cyclic cocycle, which has a more direct
geometric meaning. On $G=\PSL(2,\R)$, there is an {\em area
cocycle} (\cf \cite{Co}). This is the 2-cocycle
\begin{align*} &C:\; G\times G\to
\R\\ &C(\gamma_1,\gamma_2)=\text{(oriented) hyperbolic area of the}\\
&\hphantom{C(\gamma_1,\gamma_2)=}\text{ geodesic triangle with }\\
&\hphantom{C(\gamma_1,\gamma_2)=}\text{ vertices at}
(z_0,\gamma_1^{-1}z_0,\gamma_2 z_0),\quad z_0\in\mathbb H
\end{align*}

\smallskip

The restriction of this cocycle to a discrete subgroup
$\Gamma\subset \PSL(2,\R)$ gives the area group cocycle on
$\Gamma$. This in turn defines a cyclic 2-cocycle on $\C (\Gamma,
\sigma)$ by
\begin{equation}\label{area}
\tr_C(f_0,f_1,f_2)=\sum_{\gamma_0\gamma_1\gamma_2=1}
f_0(\gamma_0)f_1(\gamma_1)f_2(\gamma_2)C(\gamma_1,\gamma_2)
\sigma(\gamma_1,\gamma_2).
\end{equation}
Since $C$ is (polynomially) bounded, $\tr_C$ can be shown to
extend to the smooth subalgebra $\cR$.

\medskip
\subsection*{Comparison}%\hfill\medskip

Two cyclic 2-cocycles $t_1$ and $t_2$ differ by a coboundary (that
is, they define the same cyclic cohomology class) iff
$$ t_1 (a_0,a_1,a_2)- t_2 (a_0,a_1,a_2) =
 \lambda(a_0a_1, a_2)-\lambda(a_0,a_1 a_2) + \lambda(a_2 a_0,
a_1), $$ where $\lambda$ is a cyclic 1-cocycle.

\smallskip

As in \cite{CHMM}, \cite{MM1}, the difference between the
conductance cocycle $\tr_K$ and the area cocycle $\tr_C$ can be
evaluated in terms of the difference between the hyperbolic area
of a geodesic triangle and the Euclidean area of its image under
the Abel-Jacobi map. This difference can be expressed as a sum of
three terms
\begin{equation}\label{Uh}
 U(\gamma_1, \gamma_2) = h(\gamma_2^{-1}, 1) - h(\gamma_1^{-1},
\gamma_2) + h(1,\gamma_1),
\end{equation}
where each term is a difference of line integrals, one along a
geodesic segment in $\H$ and one along a straight line in the
Jacobian variety. The cocycles correspondingly differ by
$$ \tr_K (f_0,f_1,f_2)- \tr_C (f_0,f_1,f_2) =
 \sum_{\gamma_0\gamma_1\gamma_2=1} f_0(\gamma_0) f_1(\gamma_1)
f_2(\gamma_2)U(\gamma_1, \gamma_2)\sigma(\gamma_1,\gamma_2). $$
This expression can be written as $\lambda(f_0f_1,
f_2)-\lambda(f_0,f_1 f_2) + \lambda(f_2 f_0, f_1)$ where
$$\lambda(f_0,f_1)=\sum_{\gamma_0\gamma_1=1} f_0(\gamma_0)
f_1(\gamma_1) h(1,\gamma_1)\sigma(\gamma_0,\gamma_1),$$ with $h$
as in \eqref{Uh}.

\smallskip

Thus, the cocycles $\tr_K$ and $\tr_C$ differ by a coboundary.
Since they are cohomologous, $\tr_K$ and $\tr_C$ induce the same
map on $K$-theory.

\medskip
\subsection*{Values of the Hall conductance}%\hfill\medskip

The problem of deriving the values of the Hall conductance is now
reduced to computing the pairing of the area cyclic 2-cocycle with
$K$-theory. The computation is again done through an index
theorem. This time the appropriate framework is (a twisted version
of) the Connes--Moscovici higher index theorem \cite{CM}. We have
the following result, \cite{MM1}:

\begin{thm}\label{HallcondThm}
The values of the Hall conductance are given by the twisted higher
index formula
\begin{equation}\label{higherInd}
 \Ind_{c,\Gamma,\sigma}(\dirac_\cE^+\otimes \nabla)=
\frac{1}{2\pi \# G} \int_{\Sigma_{g'}} \hat A\,\, \tr(e^{R_\cE})
e^\omega u_c,
\end{equation}
where $\omega=d\eta$ is the 2-form of the magnetic field,
$\nabla^2=i\omega$, $c$ is a cyclic cocycle $c$ and $u_c$ is its
lift, as in \cite{CM}, to a 2-form on $\Sigma_{g'}$.
\end{thm}

\smallskip

Again, since $\Sigma$ is 2-dimensional, the formula
\eqref{higherInd} reduces to just the term
\begin{equation}\label{2dhInd}
 \Ind_{c,\Gamma,\sigma}(\dirac_\cE^+\otimes
\nabla) = \frac{{\rm rank}\cE}{2\pi \#G} \int_{\Sigma_{g'}} u_c.
\end{equation}

\smallskip

Notice that, while it seems at first that in \eqref{2dhInd} all
dependence on the magnetic field has disappeared in this formula,
in fact it is still present through the orbifold vector bundle
$\cE$ that corresponds (through Baum--Connes) to the class of the
spectral projection $P_E$ in $K_0(C^*_r(\Gamma,\sigma)$, of the
Fermi level.

\smallskip

When $c$ is the area cocycle, the corresponding 2-form $u_c$ is
just the hyperbolic volume form, hence the right hand side of
\eqref{2dhInd} is computed by the Gauss--Bonnet formula
$\int_{\Sigma_{g'}} u_c = 2\pi (2g'-2)$, so that
\begin{equation}\label{EulcharInd}
\frac{{\rm rank}(\cE)}{2\pi \#G} \int_{\Sigma_{g'}} u_c= {\rm
rank}(\cE) \frac{(2g'-2)}{\#G}= - {\rm rank}(\cE)
\chi_{orb}(\Sigma) \in \Q
\end{equation}
which yields an integer multiple of the orbifold Euler
characteristic.

\smallskip

The conclusion is that, in our model, the Hall conductance takes
rational values that are integer multiples of orbifold Euler
characteristics, Rational values of the conductance
$$
\sigma_H=  \tr^K(P_F,P_F,P_F)  =  \tr^C(P_F,P_F,P_F) \in \Z
\chi_{orb}(\Sigma) .
$$

\medskip
\subsection*{Discussion of the model}%\hfill\medskip

A first important observation, in terms of physical predictions,
is that our model of FQHE predicts the existence of an {\em
absolute lower bound} on the fractional values of the Hall
conductance. The lower bound is imposed by the orbifold geometry,
and does not have an analog in other theoretical models, hence it
appears to be an excellent possible experimental test of the
validity of our theoretical model. The lower bound is obtained
from the Hurwitz theorem, which states that the maximal order of a
finite group $G$ acting by isometries on a smooth Riemann surface
$\Sigma_{g'}$ is $\# G = 84(g'-1)$. This imposes the constraint on
the possible quantum Hall fractions:
$$ \phi \geq \frac{2(g'-1)}{84(g'-1)} = \frac{1}{42}. $$
The lower bound is realized by $1/42 =
-\chi_{orb}(\Sigma(0;2,3,7))$.

\smallskip

A key advantage of our hyperbolic model is that it treats the FQHE
within the same framework developed by Bellissard et al. for the
IQHE, with hyperbolic geometry replacing Euclidean geometry, to
account for the effect of electron correlation, while remaining
formally within a single particle model.

\smallskip

The fractions for the Hall conductance that we get are obtained
from an equivariant index theorem and are thus {\em topological}
in nature. Consequently, the Hall conductance is seen to be stable
under small deformations of the Hamiltonian. Thus, this model can
be generalized to systems with disorder as in \cite{CHM}, and then
the hypothesis that the Fermi level is in a spectral gap of the
Hamiltonian can be relaxed to the assumption that it is in a gap
of extended states. This is a necessary step in order to establish
the presence of plateaux.

\smallskip

In fact, this solves the apparent paradox that we still have a
FQHE, even though the Hamiltonian $H_{\sigma, V}$ may not have any
spectral gaps. The reason is that, as explained in \cite{CHM}, the
domains of the cyclic 2-cocycles $\tr_{C}$ and $\tr_{K}$ are in
fact larger than the smooth subalgebra $\cR$. More precisely,
there is a $*$-subalgebra $\mathcal A$ such that $ \cR \subset
\mathcal A \subset \cU(\Gamma, \sigma) $ and $\mathcal A$ is
contained  in the domains of $\tr_{C}$ and  $\tr_{K}$. $\mathcal
A$ is closed under the Besov space functional calculus, and the
spectral projections $P_E$ of the Hamiltonian $H_{\sigma, V}$ that
lie in ${\mathcal A}$ are called {\em gaps in extended states}.
They include all the spectral  projections onto gaps in the energy
spectrum, but contain many more spectral projections. In
particular, even though the Hamiltonian $H_{\sigma, V}$ may not
have any spectral gaps, it may still have {\em gaps in extended
states}. The results extend in a straightforward way to the case
with disorder, where one allows the potential $V$ to be random,
\cf \cite{CHM}.

\smallskip

Let us discuss the comparison with experimental data on the
quantum Hall effect. Our model recovers the observed fractions
(including the elusive $1/2$). Table 1 below illustrates how low
genus orbifolds with a small number of cone points are sufficient
to recover many observed fractions. In this first table, we
consider experimentally observed fractions, which we recover in
our model. Notice how fractions like 1/3, 2/5, 2/3, which
experimentally appear with a wider and more clearly marked
plateau, also correspond to the fractions realized by a larger
number of orbifolds (we only checked the number of solutions for
small values $\nu_j \leq 20$, $n=3$, $g=0$, and $\phi <1$). These
observations should be compared with the experimental data, \cf
\eg \cite{Stormer2} \cite{Chakra}.

\smallskip

Regarding the varying width of the plateaux, what appears
promising in Table 1 is the fact that the fractions that are more
easily observed experimentally, \ie those that appear with a
larger and more clearly marked plateau (\cf \eg \cite{Stormer2},
\cite{Chakra}), also correspond to orbifold Euler characteristics
that are realized by a large number of orbifolds. We can derive a
corresponding qualitative graph of the widths, to
be compared with the experimental ones. %see Figure \ref{Figwidth}.
Table 2 shows how to obtain some experimentally observed fractions
with $\phi>1$ (without counting multiplicities).

\smallskip

The main limitation of our model is that it seems to predict too
many fractions, which at present do not seem to correspond to
experimentally observed values. To our knowledge, however, this is
also a limitation in the other theoretical models available in the
literature. Another serious limitation is the fact that this model
does not explain why even denominator fractions are more difficult
to observe than odd ones. In fact, even for small number of cone
points and low genus, one obtains a large number of orbifold Euler
characteristics with even denominator, which are not justified
experimentally. On the occurrence of even denominators in the
fractional quantum Hall effect experiments, \cf \cite{WEST}
\cite{SSESS} \cite{EBPWH}. Table 3 provide a list of odd and even
denominator fractions predicted by our model, using genus zero
orbifolds with three cone points.

\medskip
\subsection*{Questions and directions}%\hfill\medskip

We have discussed the transition from classical Bloch theory to
noncommutative Bloch theory, as effect of the presence of a
magnetic field. In particular, we have seen that the Brillouin
zone becomes a noncommutative space. It would be interesting to
investigate, using this point of view based on noncommutative
geometry, what happens to the algebro-geometric theory of Fermi
curves and periods.
Another natural question related to the results discussed here is
whether a Chern--Simons approach to the fractional quantum Hall
effect may give a different justification for the presence of the
orbifolds $\Sigma(g;\nu_1,\ldots,\nu_n)$. In fact, these and their
symmetric products appear as spaces of vortices in Chern--Simons
(or Seiberg--Witten) theory.

\subsection*{Tables 1 and 2: experimental fractions}%\hfill\medskip
\hfill

{\small
\begin{tabular}{|c||c|||c||c|} \hline
experimental &   $g=0$ $n=3$  &  experimental &   $g=0$ $n=3$
\\ \hline  \hline
$1/3$   & $\Sigma(0;3,6,6)$  &    $2/5$   & $\Sigma(0;5,5,5)$ \\
        & $\Sigma(0;4,4,6)$  &     & $\Sigma(0;4,4,10)$ \\
    & $\Sigma(0;3,4,12)$  &     & $\Sigma(0;3,6,10)$ \\
    & $\Sigma(0;2,12,12)$  &     & $\Sigma(0;3,6,10)$ \\
    & $\Sigma(0;2,10,15)$  &         & $\Sigma(0;3,5,15)$ \\
    & $\Sigma(0;2,9,18)$  & & $\Sigma(0;2,20,20)$ \\ \hline
$2/3$   & $\Sigma(0;9,9,9)$  & $3/5$   & $\Sigma(0;5,10,10)$   \\
    & $\Sigma(0;8,8,12)$  &     & $\Sigma(0;6,6,15)$ \\
    & $\Sigma(0;6,12,12)$  &     & $\Sigma(0;4,12,15)$ \\
    & $\Sigma(0;6,10,15)$ &     & $\Sigma(0;4,10,20)$ \\
    & $\Sigma(0;6,9,18)$ & & \\
    & $\Sigma(0;5,15,15)$ & & \\
    & $\Sigma(0;5,12,20)$ & & \\ \hline
$4/9$   & $\Sigma(0;3,9,9)$  & $5/9$   & $\Sigma(0;6,6,9)$  \\
    & $\Sigma(0;4,4,18)$  &     & $\Sigma(0;4,9,12)$ \\
    & $\Sigma(0;3,6,18)$ &     & $\Sigma(0;3,18,18)$ \\  \hline
$4/5$   & $\Sigma(0;15,15,15)$  & $3/7$   & $\Sigma(0;4,4,14)$  \\
    & $\Sigma(0;12,15,20)$  & & $\Sigma(0;3,6,14)$  \\
    & $\Sigma(0;10,20,20)$  & & \\ \hline
$4/7$   & $\Sigma(0;7,7,7)$ & $5/7$   & $\Sigma(0;7,14,14)$
\\ \hline
\end{tabular}
}

%\bigskip

{\small
\begin{tabular}{|c||c|} \hline
experimental &   $g=0$ or $g=1$ \\ \hline \hline $8/5$ &
$\Sigma(0;2,4,4,5,5)$ \\ \hline $11/7$ & $\Sigma(0;2,2,7,7,7)$ \\
\hline $14/9$ & $\Sigma(1;3,9)$ \\ \hline $4/3$ & $\Sigma(1;3,3)$
\\ \hline $7/5$ & $\Sigma(0;5,5,10,10)$  \\ \hline $10/7$ &
$\Sigma(0;7,7,7,7)$ \\ \hline $13/9$ & $\Sigma(0;6,6,9,9)$ \\
\hline $5/2$ & $\Sigma(1;6,6,6)$  \\ \hline
\end{tabular}
}

%\bigskip
%\newpage
\smallskip
\subsection*{Table 3: predicted fractions}%\hfill\medskip
\hfill

{\small
\begin{tabular}{|c||c|||c||c|} \hline
odd & $g=0$ $n=3$  &  even & $g=0$ $n=3$ \\
\hline \hline
$8/15$  &  $\Sigma(0;5,6,10)$  $\Sigma(0;5,5,15)$  & $1/2$   & $\Sigma(0;6,6,6)$ $\Sigma(0;5,5,10)$ \\
    &  $\Sigma(0;4,6,20)$ $\Sigma(0;3,15,15)$ &     & $\Sigma(0;4,8,8)$ $\Sigma(0;4,6,12)$ \\
    &  $\Sigma(0;3,12,20)$ & & $\Sigma(0;4,5,20)$ $\Sigma(0;3,12,12)$ \\
    & & & $\Sigma(0;3,10,15)$ $\Sigma(0;3,9,18)$ \\ \hline
$7/9$   &  $\Sigma(0;12,12,18)$ $\Sigma(0;10,15,18)$ & $1/4$   & $\Sigma(0;4,4,4)$ $\Sigma(0;3,4,6)$ \\
    &  $\Sigma(0;9,18,18)$ &      & $\Sigma(0;3,3,12)$ $\Sigma(0;2,8,8)$  \\
    & &     & $\Sigma(0;3,3,12)$ $\Sigma(0;2,8,8)$ \\
    & &      & $\Sigma(0;2,6,12)$ $\Sigma(0;2,5,20)$ \\ \hline
$11/21$ &  $\Sigma(0;6,6,7)$ $\Sigma(0;4,7,12)$ & $7/12$  & $\Sigma(0;6,8,8)$ $\Sigma(0;6,6,12)$ \\
    &  $\Sigma(0;3,14,14)$  &    & $\Sigma(0;5,6,20)$ $\Sigma(0;4,12,12)$ \\
    & &  & $\Sigma(0;4,10,15)$ $\Sigma(0;4,9,18)$ \\ \hline
$16/21$ &  $\Sigma(0;12,12,14)$ $\Sigma(0;10,14,15)$ & & \\
    &  $\Sigma(0;9,14,18)$ & & \\ \hline
$11/15$ &  $\Sigma(0;10,10,15)$ $\Sigma(0;10,12,12)$ & &   \\
    &  $\Sigma(0;9,10,18)$ $\Sigma(0;6,20,20)$ & & \\ \hline
\end{tabular} }


\bigskip


\begin{thebibliography}{99}

\bibitem{ASS} J.~Avron, R.~Seiler, B.~Simon, {\em Charge
deficiency, charge transport and comparison of dimensions}, Comm.
Math. Phys. Vol.159 (1994), no. 2, 399--422.

\bibitem{ASY} J.~Avron, R.~Seiler, I.~Yaffe,
{\em Adiabatic theorems and applications to the integer quantum
Hall effect}, Commun. Math. Phys. Vol.110 (1987) 33--49.

\bibitem{Bell} J.~Bellissard, A.~van Elst, H.~Schulz-Baldes, {\em The
noncommutative geometry of the quantum Hall effect}, J.Math.Phys.
35 (1994) 5373--5451.

\bibitem{BellC} J.~Bellissard, {\em The noncommutative geometry of
aperiodic solids}, in ``Geometric and topological methods for
quantum field theory (Villa de Leyva, 2001)'',  86--156, World
Scientific, 2003.

\bibitem{CHMM} A.~Carey, K.~Hannabuss, V.~Mathai, P.~McCann,
 {\em Quantum Hall Effect on the hyperbolic plane},
Commun. Math. Physics, Vol.190, no. 3 (1998) 629--673.

\bibitem{CHM} A.~Carey, K.~Hannabuss, V.~Mathai,
 {\em Quantum Hall Effect on the
 Hyperbolic Plane in the presence of disorder},
Letters in Mathematical Physics, Vol. 47 (1999) 215--236.

\bibitem{CHM2} A.~Carey, K.~Hannabuss, V.~Mathai,
 {\em Quantum Hall effect and noncommutative geometry},
arXiv:math.OA/0008115.

\bibitem{Chakra} T.~Chakraborti, P.~Pietil\"anen,
{\em The Quantum Hall Effects}, Second Edition, Springer 1995.

\bibitem{FSdata} T.-S.~Choy, J.~Naset, J.~Chen, S.~Hershfield, and
C.~Stanton. {\em A database of fermi surface in virtual reality
modeling language (vrml)}, Bulletin of The American Physical
Society, 45(1):L36 42, 2000.

\bibitem{SSESS} R.G.Clark, R.J.Nicholas, A.Usher, C.T.Foxon,
J.J.Harris, {\em Surf.Sci.} 170 (1986) 141.

\bibitem{Connes} A.~Connes, {\em Non--commutative differential geometry},
Publ.Math. IHES, Vol.62 (1985) 257--360.

\bibitem{Co} A. Connes, {\em Noncommutative geometry}. Academic Press,
Inc., San Diego, CA, 1994.

\bibitem{CM} A.~Connes, H.~Moscovici,
 {\em Cyclic cohomology, the Novikov conjecture and hyperbolic
 groups}, Topology, Vol. 29 (1990) no. 3, 345--388.

\bibitem{MDY}  J.~Dodziuk, V.~Mathai, S.~Yates, {\em Arithmetic properties
of eigenvalues of generalized Harper operators on graphs}, arXiv
math.SP/0311315

\bibitem{EBPWH} {J.P.Eisenstein, G.S.Boebinger, L.N.Pfeiffer, K.W.West,
S.He}, {\em Phys. Rev. Lett.} 68 (1992) 1383; {S.Q. Murphy,
J.P.Eisenstein, G.S.Boebinger, L.N.Pfeiffer, K.W.West}, {\em Phys.
Rev. Lett.} 72 (1994) 728.

\bibitem{Gies} D.~Gieseker, H.~Kn\"orrer, E.~Trubowitz, {\em The
geometry of algebraic Fermi curves}, Perspectives in Mathematics,
Vol.14. Academic Press, 1993. viii+236 pp.

\bibitem{Gies2} D.~Gieseker, H.~Kn\"orrer, E.~Trubowitz, {\em An
overview of the geometry of algebraic Fermi curves}, in
``Algebraic geometry: Sundance 1988'', 19--46, Contemp. Math.
Vol.116, Amer. Math. Soc. 1991.

\bibitem{Gruber} M.~Gruber, {\em Noncommutative Bloch theory}, J.Math.Phys.
Vol.42  (2001),  no. 6, 2438--2465.

\bibitem{Hall} E.H.~Hall,
{\em On a new action of the magnet on electric currents},  Amer.
J. of Math. Vol.287, (1879) N.2.

\bibitem{Ji} R.~Ji, {\em  Smooth dense subalgebras of reduced group 
$C\sp *$-algebras, Schwartz cohomology of groups, and cyclic cohomology},
J. Funct. Anal. 107 (1992), no. 1, 1--33.

\bibitem{vKl} K.~von Klitzing, G.~Dorda, and M.~Pepper,
{\em New method for high--accuracy determination of the fine--structure
 constant based on quantized hall resistance},
Phys. Rev. Lett., Vol. 45 (1980) N.6, 494--497.


\bibitem{KMS} Y. Kordyukov, V. Mathai and M.A. Shubin,
{\em Equivalence of spectral projections in semiclassical limit and a vanishing theorem for higher traces in K-theory},
J.Reine Angew.Math. (Crelle),  Vol.581 (2005) 44 pages (to appear).

\bibitem{La} B.~Laughlin,
{\em Quantized hall conductivity in two dimensions}, Phys. Rev. B,
Vol.23 (1981) 5232.

\bibitem{MM1}  M.~Marcolli and V.~Mathai, {\em Twisted index theory on
good orbifolds, II: fractional quantum numbers}, Communications in
Mathematical Physics, Vol.217, no.1 (2001) 55--87.

\bibitem{MM2} M.~Marcolli and V.~Mathai, {\em Twisted index theory on good
orbifolds, I: noncommutative Bloch theory}, Communications in
Contemporary Mathematics, Vol.1 (1999)  553--587.

\bibitem{Mathai} V.~Mathai,
  {\em On positivity of the Kadison constant and noncommutative Bloch
  theory},
Tohoku Mathematical Publications, Vol.20 (2001)  107--124.

\bibitem{MaShu} V.~Mathai, M.~Shubin,
{\em  Semiclassical asymptotics and gaps in  the spectra of
magnetic Schr\"odinger operators},  Geometriae Dedicata, Vol. 91,
no. 1, (2002) 155--173.

\bibitem{RGPhi} R.G.M$\cup\Phi$, {\em
The fractional quantum Hall effect, Chern-Simons theory, and
integral lattices}, in ``Proceedings of the International Congress
of Mathematicians'', Vol. 1, 2 (Z'rich, 1994), 75--105,
Birkh\"auser, 1995.

\bibitem{Stormer2} H.L.~St\"ormer, {\em Advances in solid state
physics}, ed. P.Grosse, vol.24, Vieweg 1984.

\bibitem{Sun} T.~Sunada,
 {\em A discrete analogue of periodic magnetic Schr\"odinger operators},
Contemp. Math. Vol.173 (1994) 283--299.

\bibitem{Thou} D.J.~Thouless, M.~Kohmono, M.P.~Nightingale, M.~den
Nijs, {\em Quantized Hall conductance in a two-dimensional
periodic potential}, Phys. Rev. Lett. 49 (1982) N.6, 405--408.

\bibitem{WEST} R.Willett, J.P.Eisenstein, H.L.St\"ormer, D.C.Tsui,
A.C.Gossard, J.H.English, {\em Phys. Rev. Lett.} 59 (1987) 1776.

\bibitem{Xia} J.~Xia,
 {\em Geometric invariants of the quantum hall effect},
Commun. Math. Phys. Vol. 119 (1988), 29--50.

\end{thebibliography}
\end{document}